# Roles of LLMs in the Overall Mental Architecture

Ron Sun

Department of Cognitive Science

Rensselaer Polytechnic Institute

Troy, NY 12180, USA

dr.ron.sun@gmail.com

518-364-7439

## Abstract

To better understand existing LLMs, we may examine the human mental (cognitive/psychological) architecture, and its components and structures. Based on psychological, philosophical, and cognitive science literatures, it is argued that, within the human mental architecture, existing LLMs correspond well with implicit mental processes (intuition, instinct, and so on). However, beyond such implicit processes, explicit processes (with better symbolic capabilities) are also present within the human mental architecture, judging from psychological, philosophical, and cognitive science literatures. Various theoretical and empirical issues and questions in this regard are explored. Furthermore, it is argued that existing dual-process computational cognitive architectures (models of the human cognitive/psychological architecture) provide usable frameworks for fundamentally enhancing LLMs by introducing dual processes (both implicit and explicit) and, in the meantime, can also be enhanced by LLMs. The results are synergistic combinations (in several different senses simultaneously).

# Acknowledgments

The author benefited from discussions with a number of colleagues.

# 1. Introduction

Large language models (LLMs), which are usually based on the Transformer neural network model (Vaswani et al., 2017), having an enormous number of adjustable parameters, and trained on a massive amount of data, have achieved spectacular successes. These models have found their way into large-scale applications, dealing with not only texts but also images, voices, and other modalities. However, at the same time, unlike humans, existing LLMs suffer from some fundamental shortcomings, including limited abstract reasoning and planning capabilities, lack of human-like generalization, limited reliability and trustworthiness, lack of explainability, and so on. To improve such LLMs, it is reasonable that we look into, and draw inspirations from, the study of the human mind, which seems to be able to overcome these problems (at least to some extent).

In the opposite direction and complementarily, understanding and modelling of the human mind can benefit from better understanding and utilizing LLMs. Understanding of the human mind has always relied on latest technological advances: over time, from simple mechanical devices to complex electronic computers, which served both as metaphors and as inspirations (Boden, 2008). Now it may be the turn for LLMs to play a similar role.

Along such a line, do LLMs correspond, in any way, to any components of the human mind in the overall mental (cognitive/psychological) architecture of the human mind (or to the human mind as a whole)? If so, what components do they correspond to? How exactly do they correspond to these components if they do?

LLMs have many known strengths and also many known shortcomings. Do these strengths and shortcomings of theirs correspond, in any way, to characteristics of various components of the human mind? If they do, then can the overall architecture of the human mind (or its various components) suggest remedies for the shortcomings of LLMs, by drawing solution ideas from the human mind? Can the architecture of the human mind suggest the blueprint for a more complete and more powerful intelligent system than the currently existing LLMs?

In this respect, the notion of a computational cognitive architecture (CCA) needs to be reviewed, as it will play a crucial role in the ensuing discussion. This notion denotes a general, comprehensive *computational theory* (or model) of the human mind --- its structures and its mechanisms. It is mainly concerned with scientifically validated (to a certain extent) models of the human mind, that is, psychologically realistic or plausible ones. To understand something as complex as the human mind, a comprehensive framework as well as detailed mechanistic (e.g., computational) models are needed. Hence CCAs come into play, as they are both at the same time, resulting from decades of research in cognitive science. A typical CCA, as a computational theory/model, is precise (as well as broad and domain-generic). Its major aspects are validated through extensive modeling and simulation (i.e., accounting for empirical data). Developing progressively better CCAs over time leads to progressively better approximations of the human mental architecture and mechanisms, as well as providing unified explanations of many psychological facets along the way.

Against this backdrop, the main purpose of the present article is to advance the following four inter-related propositions:

1. Within the human mental (cognitive/psychological) architecture, LLMs may correspond with human intuition and human instinct (as well as other implicit mental processes as will be discussed later; Cosmides & Tooby, 1994; Sun & Wilson, 2014).
2. But, *in addition to* LLMs, explicit (symbolic) processes are also needed, in order to capture the human mind in its entirety. Hence dual processes (both intuition/instinct and explicit thinking) are needed.
3. Dual-process CCAs can help, as overarching psychological frameworks that incorporate both intuition/instinct and explicit thinking, in fundamentally enhancing LLMs with dual processes.
4. LLMs can in turn help to enhance CCAs in a number of important ways, leading to better computational models of the human mind.

These four propositions are important, because they together can lead to greatly enhanced CCAs, as tools for understanding the human mind, as well as greatly enhanced LLMs, as general-purpose intelligent systems, as will be detailed later.

Below, we first address Propositions 1 and 2, arguing for the need for both types of processes in capturing the human mind. Then, Propositions 3 and 4 are addressed, through the discussion of an enhanced CCA that incorporates LLMs in a fundamental way. A few concluding remarks then end the paper.

# 2. Propositions 1 and 2: Natural Division of Implicit and Explicit Processes

To address Proposition 1, a quicky review of some background regarding intuition and instinct is in order.

## 2.1. Intuition

First, what is intuition that LLMs are supposed to capture? According to New Oxford American Dictionary, intuition is "the ability to understand something immediately, without the need for conscious reasoning". According to Merriam-Webster Dictionary, it is "the power or faculty of attaining direct knowledge … without evident rational thought and inference". That is, intuition denotes thoughts without conscious reasoning or thinking --- quick, automatic, based presumably on contexts, patterns, and experiences (Sun & Wilson, 2014).

However, one may view intuition as a form of reasoning nevertheless: Reasoning encompasses explicit processes on the one hand and implicit processes (intuition) on the other (Sun, 1994). Intuition, as well as insight resulting from intuition, are arguably indispensable elements of reasoning, supplementing and guiding explicit reasoning (Helie & Sun, 2010). Both intuition and explicit reasoning are guided in turn by one's motives, needs, goals, and so on (Sun & Wilson, 2014)

To make this notion more concrete to get a better sense of it, we may examine some specific characteristics of human intuition, as gathered from empirical research in psychology and cognitive science. In fact, there are sizable literatures out there. For instance,

- Implicit Learning: First, intuition may develop on the basis of (repeated) experiential learning, which often occurs without conscious awareness (especially of details). For example, recognizing faces or detecting subtle correlations would be such cases (as demonstrated by, e.g., Lewicki, 1986; Reber, 1989; see also Seger, 1994 for a review).

- Implicit Knowledge: Intuition may rely on information that one does not or cannot explicitly access and articulate --- a vast mental repertoire of unconscious recognitions, distilled experiences, implicit heuristics, and so on (as shown by Roediger, 1990; Schacter, 1987; and others).

- Statistical Patterns: Intuition often relies on (roughly) statistical patterns and regularities, unconsciously detected and utilized (as shown by, e.g., Hasher & Zack, 1979; Lewicki, 1986; Nisbett et al., 1983; Reber & Allen, 1978).

- Similarity-based/Constraint-based/Associative Processes: Intuition often involves similarity between current situations and past experiences, as well as constraint satisfaction, heuristics, and associative processes (as opposed to just logic; Helie & Sun, 2010; Kahneman, 2011; Sloman, 1993; Sun, 1994; Tversky & Kahneman, 1974).

- Executive control: Intuition requires no explicit stepwise control and no extensive use of working memory (e.g., Evans, 2003; Evans & Frankish, 2009).

And so on. In all, intuition incorporates implicit learning and knowledge, associative processes, statistical patterns, similarity, and so on. It is an essential aspect of the human mind, allowing one to reach conclusions and render judgments quickly and efficiently, especially in familiar situations.

Now turning to the relationship between human intuition and computation, there have been some relevant theoretical discussions in the past. For instance, Dreyfus & Dreyfus (1986) provided extensive philosophical analysis of intuition. They proposed a five-stage model of development of intuition (with the five stages being novice, advanced beginner, competent, proficient, and expert), which is still relevant today. However, ironically, the title of their otherwise interesting book was "Mind over machine: The power of human intuition and expertise in the era of the computer". In other words, they pitted human intuition again computers. Their point was that machines (computers) cannot capture human intuition. However, it is important to note that, by computation, they meant mainly symbolic computation, which was prevalent at the time when the book was written. So, a proper rephasing of their point should be: Symbolic computation cannot capture human intuition.

Here, our claim is rather different and, in a way, complementary to theirs: Machines, namely LLMs (a different kind of computation), can indeed capture human intuition (which will be detailed later). Of course, LLMs were not available in the 1980s, so Dreyfus and Dreyfus' (1986) pessimism might be somewhat understandable.

## 2.2. Instinct

The next question is: What is instinct that LLMs are also supposed to capture? According to Merriam-Webster Dictionary, it is "behavior that is mediated by reactions below the conscious level". According to New Oxford American Dictionary, it is "an innate, typically fixed pattern of behavior in animals in response to certain stimuli". However, in the second definition, we should cross out "innate" and "in animal", in order for it to be consistent with the first definition, notwithstanding the fact that instincts are indeed largely rooted in our evolutionary history, guiding automatic responses to stimuli, in humans and animals alike.

As with intuition, to better understand instinct, we may examine specific characteristics of human instinct. Some characteristics can be garnered from empirical research in psychology, ethology, and cognitive science, which, however, largely mirror those of intuition as discussed earlier. For instance,

- Innate Reflexes: Instinct is often, though not always, innate (e.g., sucking for milk, or jumping away from snakes; Cosmides & Tooby, 1994).

- Implicit Learning: Instinct may also develop unconsciously. For instance, implicitly acquiring motor skills through experience, learning complex sequential reactions through experience, or instrumental conditioning are examples of implicit formation of instinct (see, e.g., Berry & Broadbent, 1984; Lewicki et al., 1987; Willingham et al., 1989).

- Implicit Knowledge: Instinct utilizes information that one does not or cannot explicitly access --- unconscious recognitions, experiences, action propensities, and so on (e.g., Lambert et al., 2003; Norman & Shallice, 1986; Shiffrin & Schneider, 1977).

- Statistical Patterns: Instinct often exploits (roughly) statistical patterns and regularities (e.g., Hasher & Zack, 1979; Lewicki, 1986).

- Associative/Reactive Processes: Instinct often involves associative or reactive processes to guide actions (e.g., Sun, 2002; Sun et al., 2001).

- Automaticity: Instinct is usually fast without explicit control (Shiffrin & Schneider, 1977; Sun et al., 2001; Wegner & Bargh, 1998).

And so on. Therefore, instincts are often deeply ingrained and usually operate on the basis of contexts, associations, and patterns, with little or no conscious thinking. They are crucial in shaping behavior, especially in determining quick responses, allowing one to navigate the world efficiently.

In terms of the relationship between human instinct and computation, our claim is similar to that regarding intuition: Machines (namely, LLMs) can capture instinct, just like they can capture intuition (more on this later).

## 2.3. Beyond Intuition and Instinct

Beyond intuition and instinct, what else is there in the human mind that is important? Intuition and instinct may be lumped into what one may term implicit processes (which, of course, may also include, e.g., low-level sensory and motor processes, and other processes of similar nature; see Reber, 1989; Sun, 2016). Now, beyond implicit processes, what else is needed to understand the full extent of the human mind? What is still missing?

Quite a few things. Here is a partial list to begin with. First, a symbolic processing capability with productivity, compositionality, and systematicity (i.e., more than just occasional cases of apparent symbol manipulation) is characteristic of human thoughts and languages, but not adequately captured by intuition and instinct (recall the debates in the 1980s; Fodor, 1975; Fodor & Pylyshyn, 1988). Also missing is rigorous logical reasoning (more than just some simple fragments of logic), which (educated) humans are clearly capable of (Bringsjord et al., 2023). More generally speaking, explicit human thinking, which is controlled (as opposed to being automatic), deliberate (as opposed to being associative or reactive), effortful, usually working memory intensive, and often rule-governed, are needed beyond intuition and instinct (as argued by, e.g., Evans, 2003; Kahneman, 2011; Sun, 1994). And so on and so forth.

Therefore, what this discussion seems to point to is the presence of *dual* processes in the human mind, as argued by researchers from psychology, philosophy, and other disciplines. Dual-process theories (DPTs) have been around for at least four decades, which hypothesize the co-existence of two qualitatively different types of mental processes (or systems) --- what one may term implicit vs. explicit processes, or System 1 vs. System 2 (or other similar dichotomies). For theoretical arguments and empirical evidence, see Evans (2003), Evans and Frankish (2009), Kahneman (2011), Macchi et al. (2016), Reber (1989), and Sun (1994, 2002).[1] These two types of processes are qualitatively different in nature, but they interact and often cooperate in operation (e.g., Sun et al., 2005). All these afore-listed missing aspects seem to belong to one side of the dichotomy --- explicit processes or System 2 (note that linking explicitness to symbolic processing has been argued extensively before; see, e.g., Smolensky, 1989; Sun, 1994, 2016).

It is worth pointing out that one version of DPT is particularly relevant here, which has been around since the late 1980s and was described in three books, published over the span of three decades (i.e., Sun, 1994, 2002, 2016). This version of DPT has not been based on processing speed (i.e., not based on "fast" vs. "slow" processes, as proposed by Kahneman, 2011), but based, outwardly, on behavioral differences (as revealed by psychological experiments) and phenomenological (first-person) differences, and, inwardly, on representational differences (symbolic vs. distributed), processing differences (precise vs. approximate), and so on. It captures and explains psychological empirical data on implicit learning, implicit memory, informal reasoning, as well as other tasks and domains. The two types of processes are termed implicit (including intuition and instinct) and explicit processes, respectively (drawing the implicit-explicit distinction from the psychological experiments on implicit learning and implicit memory). Furthermore, this DPT posits and emphasizes synergistic interaction of the two types of processes (as well as empirically demonstrating their synergistic interaction; Sun et al., 2001, 2005). At this point, we will not go into issues and details of this or other DPTs but will revisit this topic later.

---

[1] DPTs have been argued for since at least the 1980s (e.g., Reber, 1989; Sun, 1994) and popularized later by Kahneman (2011). Their historical roots go much deeper and may even be traced back to Kant, Heidegger, as well as William James (1890).

## 2.4. A Few Crucial Questions Regarding Intuition

Given such a DPT, some questions immediately arise concerning its implications for the Propositions 1 and 2 that were introduced earlier regarding LLMs. These questions include:

- Can LLMs adequately capture human intuition (as well as instinct and other implicit processes) as posited in the DPT?

- Is there a difference between behavior (by LLMs) exhibiting intuition and intuition itself?

- Why can LLMs capture human intuition, computationally speaking (beyond just the hypothesis of the DPT)?

- Can LLMs capture more than intuition (or generally, more than implicit processes), computationally (that is, in contrast to the DPT)?

- Are symbolic processes needed *beyond* LLMs, computationally? (That is, is the DPT correct in positing such dual processes?)

The first question above, which concerns adequacy of LLMs, can be addressed in an empirical way (as what will follow). However, even if LLMs can capture intuition, an objection frequently brought up is that one may capture intuition-related behavior, but not intuition itself. So addressing the second question is naturally of importance. Then we need to also look into computational characteristics of LLMs, as well as various philosophical speculations concerning them, to see why and how LLMs can capture intuition computationally. The last two questions are important in relation to dual-process approaches, because one (somewhat simplistic) school of thought has been that symbolic processes will eventually be fully captured by LLMs as they get better, and thus nothing else is needed beyond LLMs. To support a dual-process approach, these last two questions need to be addressed also.

These questions cover all the important aspects of the first two propositions introduced earlier. Answering them should provide an adequate defense of Propositions 1 and 2. Note that, below, in answering these questions, for the sake of lengths and for avoiding redundancy, focus will be on intuition, rather than all implicit processes. Let us address these questions one by one below.

### 2.4.1. Empirically, Can LLMs Capture Intuition?

Although focus will be on intuition, our discussion is applicable to other implicit processes (especially instinct) as well. Below we examine some examples of empirical confirmations of capturing various aspects of intuition by LLMs.

For instance, Han et al. (2023) showed that an LLM, when performing property induction tasks, captured various effects found in human performance of this task, including effects of premise typicality, premise diversity, premise-conclusion similarity, and so on. That is, humans and (at least some) LLMs showed similar inductive biases, which were notably not based on logic or any other formal systems.

Similarly, Dasgupta et al. (2022) discovered that an LLM showed content-sensitivity in reasoning in ways similar to humans: It was more accurate when the logically correct hypothesis was also believable content-wise. The work indicated that humans and LLMs similarly relied on semantic content in reasoning, rather than on logic alone. See also Saparov & He (2022) for other effects in reasoning by LLMs.

Furthermore, Trott et al. (2023) showed that, in the false belief task (which was to predict false beliefs of others), an LLM responded correctly 74.5% of the time, compared to humans' 82.7%. That is, humans and LLMs demonstrated similarly imperfect belief attribution. More generally, it has been shown that a human-like "theory of mind" can also emerge from LLMs (e.g., Gandhi et al., 2023; Kosinski, 2023).

Some research groups showed that embeddings from a text-only LLM mapped onto human visual representations and human visual intuition. For example, Marjieh et al. (2022) showed that embeddings from human scene descriptions, generated by a text-only LLM, predicted human visual scene similarity judgments (*r= 0.4-0.8*). Marjieh et al. (2023) showed that color similarity judgments from a text-only LLM (using the Hex code for color) were significantly correlated with human color similarity judgments (*r= 0.6-0.9*) (see also Abdou et al., 2021; Patel & Pavlick, 2022; etc.). The same goes for spatial representations of LLMs (Liétard et al., 2021; Patel & Pavlick, 2022). More generally, it has been shown that there is a correspondence between the representational space of LLMs trained on text only and the representational space of deep learning models trained on images only; linear mappings link these spaces (Li, Kementchedjhieva & Søgaard, 2023).

There have been many empirical studies in many different domains that show how LLMs may capture human intuition; only a few are listed here. It appears justified to claim that LLMs can capture human intuition in many cases, even though LLMs often have input of more limited modalities than humans.

There have also been some overall assessments and arguments in this regard. Bubeck et al. (2023) assessed performance of GPT-4 on a variety of tasks and concluded that it could capture "fast" (i.e., intuitive, implicit) processes, but not "slow" (i.e., deliberate, explicit) processes. Mugan (2023) also argued that LLMs captured implicit, unconscious processes. Sun (2024) reached a similar conclusion. Historically, others made similar or related points even before the advent of LLMs, with regard to neural network models in general (not specifically about LLMs, of course). For instance, Smolensky (1989), Sun (1994), and others made such arguments or assessments in the 1980s/90s.

### 2.4.2. Is There a Difference Between Behavior Exhibiting Intuition and Intuition Itself?

However, we must also address the second question asked earlier: Is there a difference between behavior exhibiting intuition and intuition itself? Some claim that LLMs can capture behavior exhibiting intuition but not intuition itself. Is there such a difference?

To address this question, we may look at an analogous question: Is there any difference between behavior exhibiting understanding and understanding per se? John Searle's Chinese room argument seems to suggest that there is (Searle, 1980, 1990). For the sake of lengths, we cannot possibly get into this messy debate that has been occurring for long in any detail. But it is worth pointing out that Searle's Chinese room cannot possibly, in reality, produce *realistic* behavior exhibiting understanding by any objective measures or tests. For example, apply the Turing test as originally specified by Turing (1950) --- It will, without question, fail the test (due to complexity, speed, and other difficulties). One also cannot easily explain this failure away by relying on the questionable distinction between competence and performance. So, the lesson seems to be: No understanding means no behavior exhibiting understanding --- There is no dissociation there at all. Likewise, there is likely no dissociation between intuition and behavior exhibiting intuition: If there is no intuition, there is likely no (reasonably broad range of) behavior exhibiting intuition; if there is (a reasonably broad range of) behavior exhibiting intuition, there is likely intuition per se.

There, of course, can be various possible objections to this simple, straightforward answer, for example, this one by Searle himself: What if there are many people in that room (e.g., they together emulate a neural network, with each person as a node, so they together are fast, complex, and so on) so that they can pass the Turing test together? However, note that to perform as well as humans, there may need to be billions of people in that "room", judging from the sizes of recent, more human-like LLMs. If they collectively pass the Turing test, then we should say that there is indeed understanding --- collective, but not individual, understanding, which is likely intuitive (at least to begin with). Of course, there may be other kinds of objections, but we should not get sidetracked here.

### 2.4.3. Computationally, Why Can LLMs Capture Intuition?

Then, why are LLMs capable of capturing intuition (and behavior exhibiting intuition), computationally speaking? Merely through text-based training for next-token prediction on a large Transformer-based neural network model? To address this question, meta-theoretically, in terms of the relationship between words (text) and the world, it is useful to quote Mollo & Millière (2023): "It is important to distinguish between the proximate and ultimate functions performed by LLMs during training and inference". A proximate function may give rise to an ultimate function, where the proximate function may be next-token prediction, but the ultimate function is capturing intuition. "There are compelling reasons to believe that even LLMs trained solely on text can extract information about the world by means of the causal interactions of the agents whose linguistic outputs are reflected in the training data" (Mollo & Millière, 2023). Someone else put it even more bluntly: "Text is generated by the real world, and therefore one can infer structures of the real world simply by the structural relationships within the text itself" (Hadeishi, 2024).

As an illustration, it has now been shown by a number of researchers that LLMs' representational space can be mapped to the perceptual and physical spaces (Pavlick, 2023; Sogaard, 2022; see also Marjieh et al., 2022, 2023, as mentioned earlier). Distributional associations and statistics from text can correspond well with the world (Mollo & Millière, 2023). Even skeptics of LLMs accept that there is a strong correlation between word co-occurrence statistics and semantic relationships (or relationships in the world) (e.g., Durt et al., 2023; Lenci, 2024, Titus, 2024); otherwise, fields of semantic analysis based on co-occurrences (such as LSA) would not have come into existence.

Going a little deeper, in terms of the relationship between words and the world, text "embeds information about patterns of interaction between humans and the world, human methods of categorizing…, and the selection of features…" (Mollo & Millière, 2023). In particular, sensorimotor (and other embodied) information is encoded in text (Louwerse, 2011), which was argued long ago even before LLMs. So, text captures human-specific ways of perceiving and acting in the world; those human-specific ways create the world as humans see and feel it. LLMs can capture those ways from text and then, on that basis, human intuition about the world.

Moreover, it has been claimed that "even our human experience of the world is heavily mediated --- no one has direct experience of referents" (Hadeishi, 2024): It is either through text or through sense data (sensory information), neither of which is direct. However, one may justifiably wonder if the experience is the same through text vs. through sensory information. The answer is that they are apparently close enough, according to many researchers; see, for example, many empirically studies of behavior and/or representation of LLMs as compared to human behavior and/or representation, cited earlier (for overviews, see also Binz & Schulz, 2023; Chang & Bergen, 2024; etc.). Distributional semantics acquired from text-based LLM training has shown strong correlations with true semantics, although there is likely no complete isomorphism. For instance, Lenci's (2024) study showed that LLMs captured ~80% of human understanding of partonomy, as well as other relationships (although there is no complete isomorphism). However, my key point is that distributional semantics likely leads to intuition (as discussed earlier), rather than explicit thinking; explicit thinking is the part that is missing from LLMs (as will be discussed later), which is likely why there is no complete isomorphism.

Beside the afore-discussed relationship between words and the world that links LLMs to human intuition, what about the relationship between words and the human mind or, relatedly, between LLMs and the human mind, which may also link LLMs to human intuition? According to Futrell (2023, 2024), "Language models succeed in part because they share information-processing constraints with humans. … the shared core task of language models and the brain: predicting upcoming input. … universals of language can be explained in terms of generic information-theoretic constraints, and that the same constraints explain language model performance …". Similarly, according to Ramscar (2024), meanings of words are understood in context and rely on probability distributions in context; human communication involves uncertainty reduction in predicting next words, which is not that different from LLMs. That is, LLMs and (some part of) the human mind may work in analogous ways, at least at a certain high level of abstraction, according to these researchers, who have studied language for long in accordance with this perspective.

However, some claim that LLMs are (supposedly) not "grounded" in the world and, as a result, they cannot have intuition. So now the question becomes: Are LLMs "grounded" enough to capture human intuition? To counter this objection, we need to delve briefly into the nuances of grounding. There have been different Kinds of grounding ("aboutness") in the literature: sensory-motor grounding (as advocated by Harnad, 1990; see also Sun, 2000) vs. referential grounding vs. relational grounding vs. communicative grounding, …, and so on (Lyre, 2024; Mollo & Millière, 2023). According to the arguments put forward by Mollo & Millière, referential grounding is the most essential form of grounding; it is about connecting representations to things in the world, "hooking" them onto specific entities and properties in the world; it is grounding sensu stricto (and everything else is just add-ons). LLMs (even text-only ones) can indeed be referentially grounded, through causal-informational relations between the world and representations (which are used to acquire meanings) and historical relations (e.g., evolution and learning, which lead to normativity of representations) (cf. the discourse on representational content in the philosophical literature; Mollo & Millière, 2023). This is accomplished in LLMs through indirect, mediated chains of those relations[2], thus imbuing LLMs with the ("ultimate") world-connecting functions.[3]

LLMs can certainly achieve more than just referential grounding. LLMs (even text-only ones) are capable of relational grounding and possibly other types of grounding as well (Lyre, 2024; Mollo & Millière, 2023). Multi-modal LLMs are additionally capable of sensory-motor grounding. However, are multi-modal or embodied LLMs *necessary* for grounding (in the strict sense of grounding)? The answer is no, according to the arguments outlined thus far; see both theoretical arguments and empirical results from the relevant literatures (e.g., Abdou et al., 2021; Lyre, 2024; Marjieh et al. 2022, 2003; Mollo & Millière, 2023; Pavlick, 2023; etc.). But a variety of sensory modalities and motor interactions help because they provide more information about the world (e.g., enabling sensory-motor grounding). There is a continuum in terms of quantity of information regarding the world that LLMs (or any systems) can access, leading to different degrees or types of grounding (Lyre, 2024; Mollo & Millière, 2023).

---

[2] The mediated chains of relations may involve, for example, RLHF, in-context learning, and even pre-training of LLMs, at some points of the chains (Mollo & Millière, 2023). Pretraining of LLMs may, by itself, be world-connecting, because text describes the world, its entities and properties, and normativity arises from training processes.

[3] Of course, there have been opposing views on this, which believe that LLMs cannot achieve grounding and understanding and do not have intrinsic meanings, because meanings cannot emerge only from statistical associations among linguistic forms (Bender & Koller 2020; Titus 2024; etc.). Some further claim that without sensory modalities or embodiment, there can be no real cognition (e.g., Chemero, 2023).

Now, to further argue for the linkage between LLMs and intuition, we can turn to examine some apparently shared mechanistic (computational) characteristics between human intuition and LLMs, to further strengthen the case that it is possible that LLMs can capture human intuition. On the one hand, human intuition has the following characteristics, as discussed earlier:

- Associative (Kahneman, 2011; Sun, 1994)
- Implicit (not conscious) (Reber, 1989; Schacter, 1987)
- Utilizing statistical patterns (Hasher & Zack, 1979)
- Often similarity/constraint-based (Helie & Sun, 2010; Sun, 1994)
- Not requiring executive control (Evans, 2003)
- Not usually following rigorous rules (Sun, 1994)
- Acquired through (often repeated) experiential learning, capturing diverse knowledge (Seger, 1994).

On the other hand, LLMs has the following clearly identifiable characteristics:

- Associative

- Exploiting statistical patterns
- Similarity-based
- Not conscious
- No step-by-step executive control (in terms of content-driven inference steps)
- Often not rigorously rule following
- Acquired through extensive, repeated experiences with massive data from diverse sources

These characteristics of LLMs mentioned above are self-evident (and/or already extensively discussed in the literature). Comparing these two sets, there is clearly a high degree of similarity between human intuition and LLMs. Therefore, it is not a stretch to stress the suitability of LLMs for capturing human intuition.[4]

Finally, it is worth mentioning that intuition (implicit judgment, implicit understanding, and the like) may "weakly emerge" (i.e., unexpectedly but not impossibly) from next-token predictions in training of LLMs, in accordance with the philosophical notion of "weak emergence" (Chalmer, 2006), as opposed to either direct derivation or "strong" (i.e., not deducible in principle) emergence, due to all these factors discussed so far. This notion is relevant to the earlier notion of ultimate function: Weak emergence gives rise to the ultimate function of intuition.

---

[4] Likewise, LLMs are also suitable for capturing human instinct (although it is not possible to get into details here). For instance, Chen et al. (2021) showed how Transformer models can learn how to act through a reinforcement learning setting. Relevant issues, such as proximate vs. ultimate functions, grounding, and the relationship between language and the mind, can be addressed in a way similar to how one may address these same issues with regard to intuition.

### 2.4.4. Computationally, Can LLMs Capture More Than Intuition?

Assuming LLMs can capture intuition, now the question is whether LLMs can capture, computationally, more than intuition, for example, rigorous logical reasoning or, more generally, systematic symbolic processing. Currently, the answer seems to be no (see, e.g., Bubeck et al., 2023; Chang & Bergen, 2024; Lenci, 2024). But eventually, the answer is maybe. This is because, as they get bigger, LLMs are getting better: better reasoning, fewer errors, less hallucination, and so on (as demonstrated by researchers, e.g., Chang & Bergen, 2024; Hagendorff et al., 2023; Han et al., 2024; Sartori & Orru, 2023; etc.). So, maybe eventually quantity might make up for quality: A larger number of parameters perhaps could lead to a qualitatively different outcome. This is therefore an open question. As an analogy and a cautionary tale, for example, Hubert Dreyfus predicted that computers would never defeat best human chess players, because computers would not have the power of human intuition (Dreyfus & Dreyfus, 1986). DeepBlue showed that he was wrong: Brute force raw computing power (with some knowledge engineering) may substitute for human intuition.

### 2.4.5. Why Are Symbolic Processes Needed Beyond LLMs, Computationally?

Proposition 2 claims that, *in addition to* LLMs, symbolic processes are also needed. Now we come to a crucial question: Why should LLMs not be the only thing that is needed to fully capture the human mind?

Within the context of identifying the limitations of LLMs, Lenci (2024) proposed the distinction between simple associations and formal reasoning and argued that LLMs were good at the former but not the latter. Lenci's proposal immediately brings to mind William James' (1890) distinction between "empirical thinking" and "true reasoning", which was a construct well argued for and time tested. Another idea that also comes to mind is the distinction between implicit and explicit mental processes from psychology and cognitive science, that is, the DPTs that was reviewed earlier, which suggest that we need two qualitatively different types of processes in order to fully capture the capabilities of the human mind (e.g., Reber, 1989; Seger, 1994; Sun, 1994).

In that light, what is conspicuously absent in current LLMs includes rigorous following of inference rules, procedures, and so on (as demonstrated empirically by Betz et al., 2021; Dasgupta et al., 2022; Saparov & He, 2022; etc.).[5] Another missing capability, as argued by Lenci (2024), is "true theory" (in which entities are linked by structured relations, not just flat embedding vectors), which is needed for human-like reasoning beyond LLMs. There are also other aspects (as discussed before) that can complement intuition (but not replacing intuition). All of these missing capabilities belong to one side of the dual processes, either called explicit processes (as opposed to implicit processes), System 2 (as opposed to System 1), or reason (as opposed to intuition). Evidently, this discussion of LLMs mirrors the discussion of human intuition/instinct in Section 2.3, again underlining their correspondence.

These missing capabilities require more powerful symbolic processing than what LLMs can provide currently. The full extent of these capabilities is apparently beyond neural network models based on embeddings (which, known as distributed representations in the 1980s/90s, have been investigated extensively for long) and beyond associative learning (at least so far at this point, despite decades of research on enhancing symbolic processing within neural networks). So, we need to go beyond current neural network models, including Transformers, in achieving these capabilities that are essential to human cognition.

---

[5] In empirical studies, LLMs, when given choices about possible inferences, often just predict the answer choice with the highest word overlap with the input question (Betz, Richardson, & Voigt, 2021). The models are also biased to predict intuitively plausible answers to logical questions regardless of logic. Although such effects are also present in humans to some extent, (educated) humans can eliminate such effects (e.g., in mathematical theorem proving).

Can LLMs currently capture symbolic processing? Yes, but only to a limited extent (as discussed before). As a comparison, human intuition can also carry out symbolic processing to a certain extent (e.g., language production, which is largely implicit, requires symbolic processing; Pinker & Prince, 1988). So, they correspond well with each other. Can LLMs *eventually* capture all symbolic processing? Maybe --- Quantity might have a quality of its own, as discussed before, but at least not now. However, above all and most importantly, even if LLMs could capture full symbolic processing someday, I would still argue that all symbolic processes should not be forced (e.g., trained) into LLMs.

This point can be argued from two different perspectives, aside from the usual arguments from the DPTs (e.g., Sun, 1994). First, from a third-person perspective, we obviously would like to have the following properties in any system that we create: transparency, trackability, controllability, explainability, and so on. But usual LLMs do not have these properties, so an explicit symbolic system separate from but interacting with LLMs would be desirable. On the other hand, from a first-person (introspective or psychological) perspective, a system would need various crucial psychological properties in order to fully function as the human mind would, including metacognitive monitoring/control, flexibility in reasoning/action (without extensive re-training), true one-shot learning (not just prompt engineering), and so on, all of which, again, LLMs cannot provide. So, something else --- for instance, a symbolic system on top of LLMs --- is needed. Note that this bipartite argument also serves to counter the not-so-uncommon claim regarding DPTs that dual processes (implicit and explicit) might reside within the same system (e.g., an LLM).

# 3. Propositions 3 and 4: Synergistic Combination of LLMs and CCAs

Having addressed the first two propositions, we now turn to the last two propositions. To recap, on the basis of the first two propositions, we need to further show that:

1. Dual-process CCAs, as an overarching framework that incorporates both implicit and explicit processes, can help LLMs, as they can readily enhance LLMs with symbolic capabilities and deal with the interaction between LLMs and symbolic components.
2. LLMs can in turn help CCAs, by providing CCAs with the capability of capturing human intuition (and other implicit processes) better and more completely.

The overall goal is to harness the strengths of both LLMs and CCAs, while mitigating their weaknesses. All of these objectives above can be accomplished in a cognitively/psychologically realistic or plausible way. Given the extensive discussion of Propositions 1 and 2 earlier, to argue for these two points above, it suffices to examine an example of CCAs as proof of possibility.

Recall that there have been many DPTs out there (as cited before). Likewise, there have been many CCAs, including Soar (Rosenbloom et al., 1993), ACT-R (Anderson & Lebiere, 1998), and Clarion (Sun, 2002, 2016). But the intersection of the above two categories seems very small. Among dual-process CCAs, Clarion (Sun, 2016) stands out, as it is well developed in terms of the co-existence and interaction of the two types of processes, and it is based on the DPT described in Section 2.3. (For other possibilities, see, e.g., Booch et al., 2021.)

Note also that, although CCAs have been in development since the 1970s and meant to capture human performance in a wide range of activities, their actual capabilities have been limited, not always in keeping with technological advances. For instance, some CCAs were initially conceived in the 1970s and relied on technology available then. Though some new techniques have been incorporated, they nevertheless often seem out of date. This phenomenon suggests that CCAs may need an overhaul, in order to incorporate new technological advances more readily and more naturally (especially in a dual-process framework).

Against this background, naturally, a two-fold goal would be: enhancing LLMs with Clarion and enhancing Clarion with LLMs, aiming for their synergistic combination, in a psychologically realistic or plausible way (as the Clarion framework suggests).

### 3.1. The Original Clarion Framework

A quick introduction to Clarion is in order. Clarion consists of four major subsystems: ACS, NACS, MS, and MCS (Sun, 2016). See Figure 1. ACS stands for the action-centered subsystem, which is for dealing with actions, involving procedural (i.e., action-centered) knowledge. NACS stands for the non-action-centered subsystem, which is for reasoning and memory, involving declarative (i.e., factual) knowledge. In addition, there are the motivational subsystem (MS) for dealing with motivation, and the metacognitive subsystem (MCS) for regulating other subsystems.

Clarion centers on dual processes: Each subsystem involves both implicit and explicit processes, at the bottom and the top "level" (i.e., part) of each subsystem, respectively. Computationally, one "level" is symbolic and the other neural-network-based. The two are inter-connected: The symbolic representations at the top level, which are in the forms of "chunk" nodes (each representing a concept, defined, however, via a set of microfeatures at the bottom level) and rules connecting these nodes, are linked to corresponding neural representations at the bottom level (i.e., microfeatures). The two levels interact in this way to generate combined outcomes. Thus, in Clarion, implicit and explicit processes are separate but interacting, leading to synergistic outcomes as mentioned before (Sun et al., 2001, 2005). Among different kinds of implicit processes, intuition is captured by the bottom (i.e., implicit) level of NACS, while instinct is captured by the bottom level of ACS (Sun & Wilson, 2014).

Corresponding to characteristics of implicit processes (intuition or instinct; Section 2.1-2.2), the bottom level of Clarion is implicit, subsymbolic, associative, statistical, similarity-based, and so on. The top level of Clarion is explicit, symbolic, rule-governed, precise, and so on. These characteristics are consistent with the DPT discussed in Section 2.3.

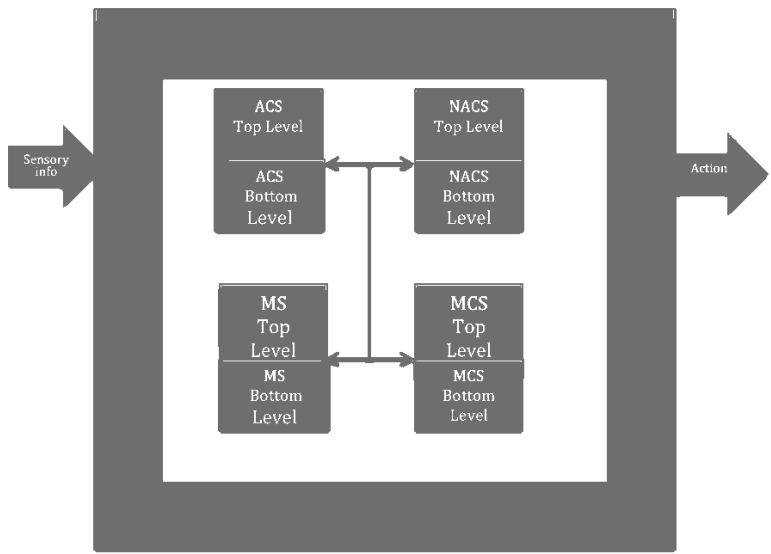

*Figure 1. The original Clarion cognitive architecture with its four major subsystems (from Sun, 2016).*

Overall, Clarion is a mechanistic, process-based psychological theory, with computational implementations, which emphasizes the interplay between implicit processes (intuition, instincts, etc.) and explicit thinking (Sun, 2016). Clarion has been well validated against psychological data, findings, and theories (Sun, 2016). By exploring this framework and its mechanisms, one gains insights into human behavior and individual differences. Clarion is also more suited to incorporating LLMs than other CCAs: Due to its foundational implicit-explicit dichotomy, LLMs can be readily incorporated into Clarion and play an essential (not peripheral) role (Romero et al., 2023; Sun, 2024).

### 3.2. The Enhanced Clarion_L Architecture

Enhancing Clarion with LLMs is straightforward; only a few relatively simple steps are needed. Essentially, at the bottom level of Clarion, LLMs are used to capture implicit processes, while symbolic explicit processes, largely unchanged, remain at the top level of Clarion.

Specifically, different implicit processes are captured separately: Intuition is captured by an LLM at the bottom level of NACS, while instinct is captured by another LLM at the bottom level of ACS. That is, two separate LLMs reside at the bottom level of these two subsystems, respectively, capturing two different kinds of implicit processes. On the other hand, MS and MCS serve as the basis of intuition and instinct (Sun & Wilson, 2014), which themselves involve implicit processes captured by LLMs. So, the overall structure of Clarion remains the same: LLMs at the bottom level for capturing implicit processes, and symbolic structures and representations at the top level for capturing explicit thinking.

This enhanced cognitive architecture is named Clarion_L, where L indicates the role of LLMs. It is Illustrated by Figure 2, where LLMs capture implicit processes.

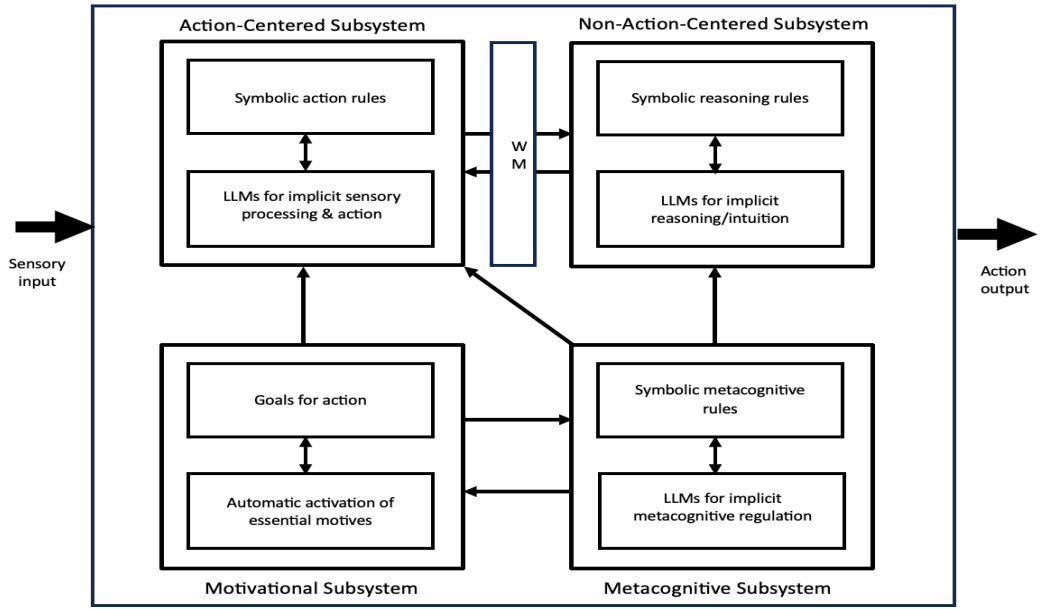

*Figure 2:* A diagram of Clarion_L. At the bottom level, LLMs are used to capture implicit processes. LLMs at the bottom level (implicit) and symbolic processes at the top level (explicit) interact and cooperate.

The central question here is: Why should we combine LLMs and symbolic processing in Clarion? Is there an advantage in combining these two according to the Clarion framework? Or one may ask a slightly more specific question: Can there be synergy resulting from combining LLMs and symbolic processing (or combining neural and symbolic processing in general) within the Clarion framework?

The answer to this last question is a definite yes. Even before the current what may be termed LLMs era, there have been plenty of demonstrations of this point, with regard to neural networks in general, such as Sun (1994), Sun et al. (2001, 2005), and so on. In the current LLM era, in the past few years, there has already been some work demonstrating this point with LLMs: for example, Romero et al. (2023), Trinh et al. (2024), and others. Notably, synergy is achieved in Clarion in a psychologically realistic or plausible way, on the basis of empirical psychological studies as well as on the basis of prior psychological validations of Clarion.

Specifically, in Clarion, implicit-explicit interactions, which lead to synergy between implicit and explicit processes, have been elucidated mechanistically (Sun, 2016), and also validated psychologically to a significant extent (through modeling and simulation of empirical psychological data; e.g., Sun et al., 2001, 2005). Such interactions can be mapped, when LLMs are incorporated into Clarion, to those between LLMs (capturing implicit processes) and explicit symbolic processes. That is, the psychologically validated dual-process structure of Clarion can be leveraged for structuring the interactions between LLMs and symbolic processes in a synergistic way.

Among other mechanisms, synergy in Clarion is achieved through (1) bottom-up and top-down learning and (2) bottom-up and top-down activation. Bottom-up learning denotes extracting explicit symbolic knowledge from intuition (or other implicit knowledge) at the bottom level as it is acquired and incorporating extracted symbolic knowledge into the top level, thus making it explicit. Top-down learning denotes assimilating explicit knowledge from the top level into intuition (or other implicit processes) at the bottom level when explicit knowledge at the top level is acquired and applied over time. Once these two types of knowledge are in place (in the top and the bottom level, respectively), Clarion utilizes both in its processing. Bottom-up activation denotes implicit processing at the bottom level gives rise to explicit outcomes by activating corresponding symbolic representations at the top level. Top-down activation denotes explicit processing at the top level triggers corresponding intuition (or other implicit processes) at the bottom level through triggering corresponding distributed representations.

Note that there is a one-to-many mapping from the top to the bottom level. That is, a unitary symbolic representation of a concept at the top level (i.e., a chunk node) maps to multiple units at the bottom level that together correspond to that concept (a distributed representation, i.e., an embedding). Therefore, once the symbolic representation is activated, the corresponding distributed representation can also be activated, and vice versa, so that the two levels can work together. (Technical details or algorithms are omitted here due to lengths.)

The two levels (implicit and explicit) work often synergistically. Synergistic effects of their interactions include that they together learn faster, perform better, and transfer better (Sun et al., 2001). Also, they together capture more accurately empirical human data in skill learning, reasoning, creative problem solving, and many other domains (Sun et al., 2001, 2005; Helie & Sun, 2010, etc.).

Within Clarion_L, the Clarion framework helps to enhance LLMs by adding enhanced symbolic capabilities on top of LLMs. More importantly, the Clarion framework helps by addressing the interaction between LLMs and symbolic processes in sophisticated (well developed) ways, so the two levels (implicit and explicit) can work together synergistically. Furthermore, the Clarion framework helps to ensure psychological realism in achieving these.

In the opposite direction, LLMs help to enhance the original Clarion, by capturing human intuition more fully, better than simpler neural networks used in CCAs before (as LLMs are much more powerful). LLMs also help to enhance Clarion by dealing with natural language input/output for it (which otherwise lacks such capabilities). Furthermore, LLMs can also deal with multi-modal input/output for Clarion (such as the visual or the auditory modality). So, taken together, the addition of LLMs enables Clarion to work in realistic environments. LLMs can elevate CCAs beyond just laboratory toys towards both theoretical tools for understanding the mind and practical tools for the real world. LLMs can better address real-world complexity, as demonstrated by many use cases seen thus far. LLMs can also lead somehow to better psychological realism of CCAs, in the sense that they generate psychologically plausible behavior in highly complex environments (beyond typically small laboratory experiments used before in validating cognitive models).

This combination is not entirely new, of course. For example, Knowles et al. (2023) proposed a CCA composed of modules made of LLMs. Kirk et al. (2024) used LLMs as a source for building up knowledge within an existing CCA. Romero et al. (2023) compared several possibilities in combining LLMs and CCAs (which include modular, agent-based, and neuro-symbolic approaches, including those utilizing Clarion). Xie et al. (2023) drew on cognitive architectures for enhancing LLMs, incorporating attention, memory, reasoning, and learning mechanisms. There are also many other agent models utilizing LLMs and CCAs (for surveys, see, e.g., Sumers et al., 2023). Among those, Clarion_L is more theory-driven and more psychologically validated.

## 4. Concluding Remarks

It is obviously important (and expedient), at the present time, to produce larger and larger LLMs, by curating and using more and more training data and running them with enormous computing power, in order to show increasingly better performance of LLMs. However, it is also crucial to explore alternative approaches and/or perspectives --- We should not put all eggs in one basket. It is equally or even more important that we continue to look into cognitive science, psychology, neuroscience, and so on. They were inspirations in the past resulting in the advances that we currently see, and they should continue to be inspirations for future advances. New, better ideas and methods may emerge that way --- from "reverse engineering" of the human mind or brain, concerning cognitive architectures, dual processes, intuition, reasoning, …., and so on. Moreover, and maybe more importantly, combinations of LLMs and CCAs can also help us to better understand, and to better capture in computational models, the human mind, likely leading to the next generation of theoretical and practical tools.

What has been emphasized in the present article is the importance of dual-process computational cognitive architectures (and consequent hybrid neuro-symbolic approaches) in addressing the limitations of currently existing models and systems. In particular, Clarion provides a theoretical foundation for the integration of CCAs and LLMs, contributing to synergistic integration of the two subdisciplines for more robust, more intelligent, and more human-like systems.

# Acknowledgments

The author benefited from discussions with a number of colleagues.

# References

Abdou, M., Kulmizev, A., Hershcovich, D., Frank, S., Pavlick, E. & Søgaard, A. (2021), Can language models encode perceptual structure without grounding? a case study in color. *Proceedings of the 25th Conference on Computational Natural Language Learning*, pp.109–132,

Anderson, J. R. & C. Lebiere, (1998). *The Atomic Components of Thought*. Lawrence Erlbaum Associates, Mahwah, NJ.

Berry, D. & D. Broadbent, (1984). On the relationship between task performance and associated verbalizable knowledge. *Quarterly Journal of Experimental Psychology*. 36A, 209-231.

Binz, M., & Schulz, E. (2023). Using cognitive psychology to understand GPT-3. *Proceedings of the National Academy of Sciences*, 120(6), e2218523120.

Boden, M. A. (2008). An evaluation of computational modeling in cognitive science. In R. Sun (Ed.), *The Cambridge handbook of computational psychology* (pp. 667–683). Cambridge: Cambridge University Press.

Booch, G., Fabiano, F., Horesh, L., Kate, K., Lenchner, J., Linck, N., Loreggia, A., Murgesan, K., Mattei, N., Rossi, F., & Srivastava, B., (2021). Thinking fast and slow in AI. *Proceedings of the AAAI Conference on Artificial Intelligence*. 15042-15046.

Bringsjord, S., Giancola, M., & Govindarajulu, N. S. (2023). Logic-Based Modeling of Cognition. In R. Sun (Ed.), *The Cambridge Handbook of Computational Cognitive Sciences* (pp. 173–209). Cambridge: Cambridge University Press.

Bubeck, S., et al. (2023). Sparks of artificial general intelligence: Early experiments with GPT-4. *arXiv:2303.12712*.

Chalmers, D. J. (2006). Strong and weak emergence. In: P. Clayton & P. Davies (eds.), *The Re-Emergence Of Emergence: The Emergentist Hypothesis From Science To Religion*. New York: Oxford University Press.

Chang, T. & B. Bergen (2024). Language model behavior: a comprehensive survey. *Computational Linguistics*. 50(1), 293–350.

Chemero, A. (2023). LLMs differ from human cognition because they are not embodied. *Nat Hum Behav,* 7, 1828–1829.

Chen, L., Lu, K., Rajeswaran, A., Lee, K., Grover, A., Laskin, M., Abbeel, P., Srinivas, A., & Mordatch, I. (2021). Decision transformer: Reinforcement learning via sequence modeling. *Advances in Neural Information Processing Systems*, 34, 15084–15097.

Cosmides, L. & J. Tooby, (1994). Beyond intuition and instinct blindness: Toward an evolutionarily rigorous cognitive science. *Cognition.* 50, 41-77.

Dasgupta, I., Lampinen, A. K., Chan, S. C., Creswell, A., Kumaran, D., McClelland, J. L., & Hill, F. (2022). Language models show human-like content effects on reasoning. *arXiv:2207.07051*.

Dreyfus, H. & Dreyfus, S. (1986). *Mind over Machine: The Power of Human Intuition and Expertise in the Era of the Compute*r. Oxford, U.K.: Blackwell.

Durt, C, Froese, T, & Fuchs, T. (2023). Large language models and the patterns of human language use: an alternative view of the relation of ai to understanding and sentience. https://philsci-archive.pitt.edu/id/eprint/22744

Dasgupta, I., Lampinen, A. K., Chan, S. C., Creswell, A., Kumaran, D., McClelland, J. L., & Hill, F. (2022). Language models show human-like content effects on reasoning. *arXiv:2207.07051*.

Evans, J. (2003). In two minds: dual-process accounts of reasoning, *Trends in Cognitive Sciences*, 7(10), 454-459.

Evans, J. & K. Frankish (eds.), (2009). *In Two Minds: Dual Processes and Beyond*. Oxford University Press, Oxford, UK.

Fodor, J. A., (1975). *The Language of Thought*, New York: Thomas Y. Crowell.

Fodor, J. A., & Pylyshyn, Z. W. (1988). Connectionism and cognitive architecture: A critical analysis. *Cognition,* 28(1-2), 3–71.

Futrell, R. (2023). Information-theoretic principles in incremental language production. *Proceedings of the National Academy of Sciences* 120(39), e2220593120.

Futrell, R. (2024). Abstract of the talk at the UQAM Summer School on Understanding LLM Understanding. June 3-14, 2024. https://skywritingspress.ca/#draft-timetable

Gandhi, K., Fränken, J. P., Gerstenberg, T., & Goodman, N. D. (2023). Understanding social reasoning in language models with language models. *arXiv:2306.15448*.

Hadeishi, M. (2024). *Connectionists Digest*, 880(4), #13.

Harnad, S. (1990). The Symbol Grounding Problem. *Physica D*, 42(1–3), 335–346.

Hasher, J. & J. Zacks, (1979). Automatic and effortful processes in memory. *Journal of Experimental Psychology: General*, 108, 356-358.

Helie, S. & R. Sun, (2010). Incubation, insight, and creative problem solving: A unified theory and a connectionist model. *Psychological Review,* 117(3), 994-1024.

Kahneman, D. (2011). *Thinking, fast and slow*. New York: Farrar, Straus and Giroux.

Kirk, J.E., R.E. Wray, & J. E. Laird, (2024). Exploiting language models as a source of knowledge for cognitive agents. arXiv:2310.06846v1

Knowles, K., M. Witbrock, G. Dobbie, & V. Yogarajan (2023). A proposal for a language model based cognitive architecture. In: C. Geib & R. Petrick (Des), *Proceedings of the 2023 AAAI Fall Symposia*. Arlington, Virginia.

Kosinski, M. (2023). Theory of mind may have spontaneously emerged in LLMs. *arXiv:2302.02083*.

Lenci, A. (2024). Talk given at the UQAM summer school on LLMs. https://skywritingspress.ca/#draft-timetable

Lewicki, P. (1986). Processing information about covariations that cannot be articulated. *Journal of Experimental Psychology: Learning, Memory, and Cognition*, 12, 135-146.

Lewicki, P., M. Czyzewska, & H. Hoffman, (1987). Unconscious acquisition of complex procedural knowledge. *Journal of Experimental Psychology: Learning, Memory and Cognition*. 13(4), 523-530.

Louwerse, M.M. (2011), Symbol Interdependency in Symbolic and Embodied Cognition. *Topics in Cognitive Science*, 3, 273-302.

Lyre, H. (2024). Understanding AI: Semantic Grounding in Large Language Models. *arXiv:2402.10992*.

Macchi, L., M. Bagassi, & R. Viale, (eds.), (2016). *Cognitive Unconscious and Human Rationality.* MIT Press, Cambridge, MA.

Marjieh, R., Sucholutsky, I., Sumers, T. R., Jacoby, N., & Griffiths, T. L. (2022). Predicting human similarity judgments using large language models. *arXiv:2202.04728*.

Marjieh, R., Sucholutsky, I., van Rijn, P., Jacoby, N., & Griffiths, T. L. (2023). Large language models predict human sensory judgments across six modalities. *arXiv:2302.01308*.

Mollo, D.C. & Millière, R. (2023), The vector grounding problem. *https://arxiv.org/abs/2304.01481*.

Mugan, J. (2023). Grounding large language models in a cognitive foundation: how to build someone we can talk to. *The Gradient.*

Nisbett, R.E., Krantz, D.H., Jepson, C., & Kunda, Z. (1983). The use of statistical heuristics in everyday inductive reasoning. *Psychological Review*, *90*, 339-363.

Norman, D., & Shallice, T. (1986). Attention to action: willed and automatic control of behavior. In: G. Schwartz & D. Shapiro (Eds.). *Consciousness and self regulation: Advances in research and theory* (Vol.4, pp.1–18). New York: Plenum.

Patel, R. & Pavlick, E. (2022), Mapping language models to grounded conceptual spaces. In *Proceedings of the International Conference on Learning Representations*.

Pavlick, E. (2023). Symbols and grounding in large language models. *Philosophical Transactions A,* 381(2251).

Pinker, S. & Prince, A. (1988). On language and connectionism: Analysis of a parallel distributed processing model of language acquisition. *Cognition*, 28(1-2), 73-193.

Ramscar, M. (2024). Talk given at the NSF Workshop on Text Production and Comprehension by Human and Artificial Intelligence. https://www.isu-pacelab.org/workshop-ai-text-production

Reber, A. (1989). Implicit learning and tacit knowledge. *Journal of Experimental Psychology: General*. 118(3), 219-235.

Reber, A. & R. Allen, (1978). Analogy and abstraction strategies in synthetic grammar learning: A functionalist interpretation. *Cognition*. 6, 189-221.

Roediger, H. (1990). Implicit memory: Retention without remembering. *American Psychologist*, 45 (9), 1043-1056.

Romero, O.J., J. Zimmerman, A. Steinfeld, & A. Tomasic (2023). Synergistic integration of large language models and cognitive architectures for robust AI: An exploratory analysis. *arXiv:2308.09830*.

Rosenbloom, P., J. Laird, & A. Newell, (1993). *The SOAR Papers: Research on Integrated Intelligence*. MIT Press, Cambridge, MA.

Saparov, A., & He, H. (2022). Language models are greedy reasoners: A systematic formal analysis of chain-of-thought. *arXiv:2210.01240*.

Schacter, D. (1987).  Implicit memory: History and current status. *Journal of Experimental Psychology: Learning, Memory, and Cognition*, 13, 501-518.

Searle, J. R. (1980). Minds, brains, and programs. *Behavioral and Brain Sciences*. 3(3), 417–457.

Searle, J. R. (1990). Is the Brain's Mind a Computer Program? *Scientific American,* 262 (1), 26-31.

Seger, C. (1994). Implicit learning. *Psychological Bulletin*, 115(2), 163-196.

Shiffrin, R.  & W. Schneider, (1977). Controlled and automatic human information processing II. *Psychological Review*, 84, 127-190.

Sloman, S. (1993).  Feature based induction.  *Cognitive Psychology*, 25, 231-280.

Sumers, T., et al. (2023). Cognitive architectures for language agents. *arXiv:2309.02427v2*

Sun, R. (1994). *Integrating Rules and Connectionism for Robust Commonsense Reasoning.* Wiley, New York.

Sun, R. (2000). Symbol grounding: A new look at an old issue. *Philosophical Psychology*, 13(3), 403-418.

Sun, R. (2002). *Duality of the Mind*.  Erlbaum, Mahwah, NJ.

Sun, R. (2016). *Anatomy of the Mind*. Oxford University Press, Oxford, UK.

Sun, R., E. Merrill, & T. Peterson, (2001). From implicit skills to explicit knowledge: A bottom- up model of skill learning. *Cognitive Science*, 25(2), 203-244.

Sun, R., P. Slusarz, & C. Terry, (2005). The interaction of the explicit and the implicit in skill learning: A dual-process approach. *Psychological Review*, 112(1), 159-192.

Sun, R. & N. Wilson, (2014). Roles of implicit processes: instinct, intuition, and personality. *Mind and Society*, 13(1), 109-134.

Titus, L. M. (2024). Does ChatGPT have semantic understanding? *Cognitive Systems Research*, 83, 1–13.

Trinh, T.H., Wu, Y., Le, Q.V. *et al.* (2024). Solving Olympiad geometry without human demonstrations. *Nature,* 625, 476–482.

Trott, S., Jones, C., Chang, T., Michaelov, J., & Bergen, B. (2023). Do large language models know what humans know? *Cognitive Science,* 47(7).

Turing, A. (1950). Computing machinery and intelligence. *Mind,* 59(236), 433–460.

Tversky, A. & Kahneman, D. (1974). Judgment under uncertainty: Heuristics and biases. *Science*, 185, 1124-1131.

Vaswani, A., Shazeer, N., Parmar, N., et al. (2017). Attention is all you need. *Advances in neural information processing systems,* 30.

Wegner, D. M., & Bargh, J. A. (1998). Control and automaticity in social life. In D. Gilbert, S. T. Fiske, & G. Lindzey (Eds.), *Handbook of Social Psychology* (4th ed, Vol.1, pp.446-496). New York: McGraw-Hill.

Willingham, D., M. Nissen, & P. Bullemer, (1989). On the development of procedural knowledge. *Journal of Experimental Psychology: Learning, Memory, and Cognition*, 15, 1047-1060.

Xie, Y., Xie, T., Lin, M., Wei, W., Li, C., Kong, B., Chen, L., Zhuo, C., Hu, B. & Li, Z., (2023). OlaGPT: Empowering LLMs with human-like problem-solving abilities. *arXiv:2305.16334*.